# Generation of Programming Exam Question and Answer Using ChatGPT Based on Prompt Engineering

Jongwook Si and Sungyoung Kim

*Abstract*—In computer science, students are encouraged to learn various programming languages such as Python, C++, and Java, equipping them with a broad range of technical skills and problem-solving capabilities. Nevertheless, the design of objective examination questions to assess students' creativity, problem-solving abilities, and domain knowledge remains a significant challenge. This paper proposes a methodology to address these challenges by leveraging prompt engineering techniques with ChatGPT. Prompt engineering is an efficient technique that optimizes the performance of language models, enabling the automatic generation of high-quality exam questions with varying types and difficulty levels, all without requiring additional fine-tuning of the model. This study applies diverse patterns and templates to generate exam questions that incorporate both theoretical and practical components, thereby facilitating a comprehensive evaluation of students' theoretical understanding and hands-on programming proficiency. A survey was conducted to validate the proposed method, and although certain areas indicated room for improvement, the overall results confirmed its significance and relevance. The generated questions and model answers exhibit quality comparable to, or even surpassing, manually crafted questions while significantly reducing the time and effort required for question preparation. This research demonstrates that automated exam question generation through prompt engineering enhances the quality and efficiency of assessment tools in education, establishing it as a valuable asset for future educational environments.

## I. INTRODUCTION

Prompt engineering has emerged as a critical technique for effectively leveraging artificial intelligence (AI) language models [1]-[3]. It refers to the process of designing and optimizing instructions or queries, known as prompts, that are provided to AI systems in order to achieve desired outcomes. With the advancement of large language models (LLMs) such as ChatGPT [4], the importance of prompt engineering has become increasingly evident. The quality of prompts directly influences the accuracy and relevance of model outputs [5]. LLMs operate as neural architectures trained to interpret human intent and contextual information [6]. Even minor variations in wording or structure can result in substantially different outputs [7], which underscores the necessity of precise and deliberate prompt design.

One of the major advantages of prompt engineering is that it allows modification of the model's output format and accuracy without the need for additional fine-tuning [8]. This reduces computational cost and time, while enabling users to exert more direct control over model behavior [9], [10]. Furthermore, context-rich prompts enhance reasoning capability and improve output quality, thereby compensating for inherent limitations of current models. This is particularly valuable in domains such as legal advisory or technical documentation where reliability and precision are essential. However, the internal mechanisms of LLMs remain opaque, which makes it difficult to quantitatively evaluate the relationship between prompt design and model performance. Token-length constraints also restrict the complexity of prompts [11]. Therefore, effective prompt engineering requires iterative experimentation and qualitative assessment.

As AI becomes increasingly integrated into everyday life and professional practice, the role of prompt engineering continues to expand [12]-[15]. Mastery of this skill contributes to productivity improvements at both the individual and organizational levels, and it has become recognized as a core competency within AI-driven industries [16]. This indicates the necessity of systematically exploring prompt engineering in educational as well as professional contexts.

Successful application of prompt engineering is guided by several fundamental principles [17]-[19]. Prompts should include clear and detailed instructions, while avoiding vague or overly general expressions. They should explicitly define task objectives and, when needed, provide supplementary context or references to improve accuracy and relevance. In addition, specifying structured output formats and styles ensures consistency and high quality [20].

ChatGPT [4] demonstrates strong performance in understanding context and predicting subsequent sentences, but its use is constrained by token limits. Another limitation is the issue of hallucination [21], where the model produces content that appears factual but is inaccurate or fabricated. This is particularly problematic when users lack sufficient domain expertise to validate outputs. In programming contexts [22], however, hallucination poses less risk because generated code can be executed and verified. For this reason, ChatGPT [4] is widely applicable across various domains [23], with

Jongwook Si is with the Department of Computer AI Convergence Engineering, 39177, Gumi, Gyeong-buk, Republic of Korea (email : jwsi425@kumoh.ac.kr)

Sungyoung kim is with the Department of Computer Engineering, 39177, Gumi, Gyeong-buk, Republic of Korea (email : sykim@kumoh.ac.kr)

especially strong potential in the field of education [24].

Recently, computer programming education has become an essential component not only in secondary and high schools but also in higher education. Traditionally, C and C++ served as foundational programming languages, yet Python has rapidly gained popularity due to its simplicity and intuitive syntax. However, the widespread adoption of language models has led students to rely on AI-generated code without acquiring a fundamental understanding of core programming concepts, which undermines the development of creative thinking and problem-solving skills. University examinations are primarily designed to assess students' theoretical knowledge and creativity, but the manual construction of high-quality exam questions is time-intensive and heavily dependent on the instructor's expertise [25]. In addition, requiring students to write complete programs from scratch presents grading challenges and fails to adequately capture their theoretical understanding. Consequently, there is a growing need to evaluate learners' overall competencies through diverse types of questions.

Programming education must cultivate creative thinking and problem-solving abilities; therefore, an assessment method that balances theoretical knowledge and practical experience is required. Nevertheless, traditional educational systems have not sufficiently supported such an approach. This study proposes a methodology for the automated generation of high-quality, creative, and diverse programming exam questions by employing ChatGPT [4] and prompt engineering techniques. To achieve this, templates encompassing both theoretical and practical problems are designed, enabling the generation of questions with varying levels of difficulty and formats. Through this approach, learners' knowledge and practical skills can be evaluated comprehensively. This methodology aims to overcome existing limitations in educational assessment and to explore the practical potential of AI-driven automated exam question generation.

The methodological procedure of this study consists of three main stages. First, templates were designed to generate various types of programming exam questions using prompt engineering techniques. In this process, question type classification, rules for generating questions and answers, and design principles based on educational theories (constructivism and Bloom's taxonomy) were incorporated. Second, the designed templates were applied to automatically generate diverse questions across multiple programming languages, including Python, C++, and Java, and the resulting items were systematically collected and organized. Third, a survey was conducted with undergraduate students majoring in programming to empirically evaluate the usefulness and educational effectiveness of the proposed problem-generation method. Participants were categorized according to their programming experience level (beginner, intermediate, advanced), which served as a quasi-control group to enable comparison of responses and identification of group-specific differences. The empirical evaluation involving participants was carried out in the third stage of the study, where students directly responded to the automatically generated questions or evaluated the perceived quality and clarity of the items. The data collected through this process were interpreted using descriptive statistics and subgroup comparisons.

Section 2 provides an overview of the types of problems that can be designed for programming courses. Section 3 introduces the templates developed through the application of prompt engineering techniques, detailing their structure and functionality. Section 4 presents an analysis of the generated results, evaluating their quality and relevance to educational objectives. The study concludes in Section 5 with a discussion of future research directions, highlighting potential improvements and extensions of the proposed methodology.

## II. BACKGROUND IN PROGRAMMING COURSES

### A. Multiple-Choice Questions

In programming education, multiple-choice questions have been shown to be an effective method for assessing students' foundational theoretical knowledge and coding skills [26]. Multiple-choice questions require students to make clear decisions within a limited timeframe, enabling educators to evaluate large cohorts efficiently. Moreover, multiple-choice questions allow instructors to diagnose students' learning status comprehensively by assessing their understanding of specific concepts, algorithmic thinking, and problem-solving skills. From the standpoint of higher-order thinking frameworks, well-designed multiple-choice questions may promote analysis, evaluation, and application rather than mere recall [27]. Therefore, when properly constructed, multiple-choice questions can encourage students to trace code execution processes and consider diverse scenarios, thereby fostering higher-level cognitive engagement [28].

Multiple-choice questions in programming education can be categorized into ten types, each targeting specific cognitive and practical skills. The first type involves selecting the correct output or identifying errors in code, promoting code tracing and debugging abilities. The second type asks for multiple correct answers about code behavior, encouraging students to consider diverse execution paths. The third type requires arranging code or algorithm steps in order, reinforcing logical and procedural thinking. The fourth type involves filling in missing code elements, supporting syntax recall and functional understanding. The fifth type focuses on choosing the best algorithm for a problem, fostering applied problem-solving skills. The sixth type matches programming terms with definitions, ensuring conceptual clarity. The seventh type involves assembling a complete program from fragments, enhancing understanding of modularity and structure. The eighth type requires explaining a program's output, developing analytical skills. The ninth type asks for descriptions that match code behavior, assessing theoretical understanding. The tenth type evaluates the ability to predict changes in program behavior following code modifications. These question types together assess a wide range of knowledge, from syntax and logic to higher-order thinking and real-world application.

## B. Short-Answer Questions

Short-answer questions consist of six categories. The first type asks students to trace the execution of a given code and predict its output, verifying understanding of core concepts such as loops, conditionals, function calls, and recursion. The second type requires explaining the purpose and behavior of a program and filling in missing code, strengthening both code analysis and implementation skills. The third type presents data visualizations such as tables or graphs and asks students to write code to process the data, thereby integrating data interpretation with programming practice. The fourth type provides only the output and asks students to write code that produces it, fostering creativity and logical problem-solving. The fifth type requires writing a program that satisfies specific conditions, training students to analyze requirements systematically and implement them effectively. The sixth type involves writing a complete program that integrates multiple functions and requirements, comprehensively assessing advanced skills in program design, implementation, and function integration. These short-answer questions allow learners to demonstrate not only programming syntax but also logical reasoning and systematic problem-solving skills.

## C. Essay Questions

Essay questions are divided into three categories. The first type requires defining specific programming terms and explaining their meaning and applications, assessing learners' conceptual understanding and ability to link theory to practice. The second type involves analyzing code errors, describing their causes, and proposing solutions, which evaluates learners' debugging skills and logical reasoning. The third type asks students to analyze inefficiencies or flaws in given code and propose improvements, thereby developing skills in optimization and refactoring while encouraging consideration of performance, readability, and maintainability.

Taken together, multiple-choice, short-answer, and essay questions reflect learning objectives that range from knowledge acquisition to analysis, evaluation, and creative problem-solving. Thus, the use of prompt engineering to generate such diverse question types is not limited to automation of assessments but serves as an educational mechanism that enhances learners' higher-order thinking skills and practical programming competence in a balanced and systematic way.

## III. PROPOSED TEMPLATES USING PROMPT ENGINEERING

Prompt engineering is a powerful method for creating reliable and effective exam questions for programming courses. By leveraging the capabilities of large language models such as ChatGPT, educators can design high-quality questions that align with their educational goals. This section explores key prompt engineering techniques that can enhance the process of generating exam questions, providing detailed explanations and their potential benefits and limitations. Figure 1 illustrates the overall flow of the proposed method.

## A. Role Assignment

Role assignment involves specifying a particular role or profession within the prompt to guide the language model toward generating responses that reflect the expertise of that role. For example, assigning roles such as "programming language expert," "computer science professor," or "software engineer" allows the model to generate questions that are more specialized and contextually relevant. The purpose of this technique is to provide the model with contextual cues that help it better understand the task or question.

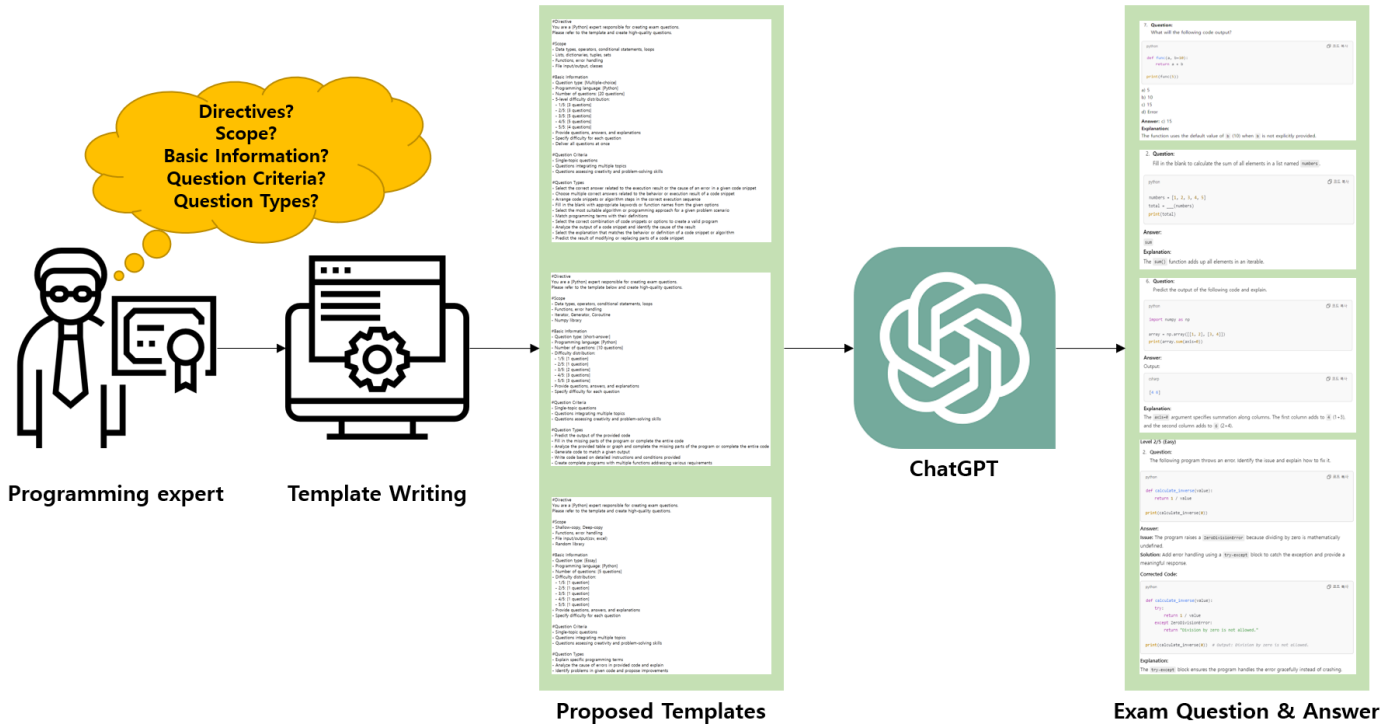

**Fig. 1.** Overall Flowchart of the Proposed Method.

By assigning a role, the model is more likely to reference relevant knowledge and generate advanced questions and answers. For instance, if the role of "programming expert" is assigned, and the prompt requests, "Create multiple-choice questions about Python list comprehensions", the resulting questions are likely to reflect an in-depth understanding of the topic, tailored to an academic context.

However, a notable limitation of this technique is the risk of hallucination, where the model may generate information that is inaccurate or inconsistent with the assigned role. Therefore, careful validation and review of the generated questions are essential to ensure their quality and accuracy. Despite this limitation, role assignment remains an attractive technique for generating high-quality, domain-specific questions when properly monitored.

*B. Few-shot Learning*

Few-shot learning involves providing a few examples in the prompt to enhance the model's ability to generate more relevant and contextually rich outputs. Unlike zero-shot learning, where no examples are given, few-shot learning uses two or more examples to help the model infer the desired structure or context of the output.

In the context of programming exams, this technique is particularly effective for creating diverse and high-quality questions. By including examples of question types, such as multiple-choice or fill-in-the-blank, the model can better understand the format and requirements of the desired output. For instance, educators can provide a sample multiple-choice question and a sample short-answer question in the prompt, which helps the model generate additional questions that are varied yet relevant.

Incorporating few-shot learning also allows for the inclusion of specific details, such as the question types discussed in Section 2. For example, providing a context that outlines the range of question types, difficulty levels, and topics ensures that the model generates questions aligned with the educational objectives. This technique enables the generation of questions that are more nuanced and comprehensive compared to zero-shot prompts.

By offering multiple examples, few-shot learning significantly improves the quality and diversity of generated questions, making it an invaluable tool for educators who aim to evaluate students' critical thinking and problem-solving abilities effectively.

*C. Fukatsu Prompting*

Fukatsu prompting is a structured approach that involves designing prompts with a clear format and template to ensure consistent and high-quality outputs [30]. This technique incorporates various components, such as directives, scope, basic information, question criteria, and question types, into a single cohesive prompt template.

The template proposed in this study consists of the following components:

- #Directives: This section specifies the role and task. For example, it may include roles like "programming

expert", "programming professor" and tasks such as "generate advanced algorithm questions for a Python exam".

- #Scope: This section defines the scope of the exam questions, such as Python data structures, algorithms, or object-oriented programming.
- #Basic Information: This includes details like question format (multiple-choice, short answer, or essay), the programming language, the number of questions, and difficulty levels (e.g., 1/5~ 5/5).
- #Question Criteria: This section outlines the content and nature of the problems to be generated, such as whether they focus on single-topic problems, multi-part problems, or creativity and problem-solving skills.
- #Question Types: This specifies the types of questions to be included, such as predicting code output, filling in missing code, analyzing errors, or combining multiple concepts.

By using this template, educators can provide all the necessary details in a structured manner, ensuring the model generates reliable and high-quality exam questions. For instance, the role assignment under #Directives helps establish the model's perspective, while the inclusion of scope and question criteria ensures the generated questions align with the desired learning objectives. Additionally, specifying the question types from Section 2 allows for greater control over the variety and format of the questions.

This structured approach ensures that the model generate questions that are not only high-quality but also tailored to the specific requirements of the course. It allows for a streamlined workflow, enabling educators to efficiently generate a wide range of exam questions with detailed solutions.

Fig. 2 represents the proposed prompt template for multiple-choice questions, Fig. 3 illustrates the template for short-answer questions, and Fig. 4 depicts the template for essay questions, with the scope and basic information included as examples for demonstration purposes. The question types include all those mentioned in Section 2.

The multiple-choice template is designed as a tool to quickly and effectively evaluate students' understanding of basic concepts and logical reasoning. The questions are structured with clear options, enabling students to select the most appropriate answer for a given scenario. The primary feature of this template is the inclusion of unambiguous choices, maintaining clarity while encompassing a wide range of difficulty levels, from beginner to advanced, to allow comprehensive assessment. The question types include predicting code execution results, arranging code snippets in the correct sequence, and filling in keywords. This template is particularly useful for assessing students' understanding of concepts, code analysis skills, and error identification in a straightforward yet effective manner.

The short-answer template is designed to encourage students to articulate their problem-solving process and demonstrate creativity. This template goes beyond requiring a simple correct answer, prompting students to solve problems using logical and structured approaches. The questions provide clear conditions and guidelines, enabling students to write or analyze code based on the given context. Easy questions may involve predicting simple output values, while more challenging questions might require implementing complex logic. Key question types include filling in code blanks, writing code based on analyzed outputs, and completing programs by analyzing tables or graphs. The short-answer template is suitable for deeply evaluating students' problem-solving abilities, data processing skills, and logic design competencies.

```
#Directive
You are a [Python] expert responsible for creating exam questions.
Please refer to the template and create high-quality questions.

#Scope
  - Data types, operators, conditional statements, loops
  - Lists, dictionaries, tuples, sets
  - Functions, error handling
  - File input/output, classes

#Basic Information
  - Question type: [Multiple-choice]
  - Programming language: [Python]
  - Number of questions: [20 questions]
  - 5-level difficulty distribution:
    - 1/5: [3 questions]
    - 2/5: [3 questions]
    - 3/5: [5 questions]
    - 4/5: [5 questions]
    - 5/5: [4 questions]
  - Provide questions, answers, and explanations
  - Specify difficulty for each question
  - Deliver all questions at once

#Question Criteria
  - Single-topic questions
  - Questions integrating multiple topics
  - Questions assessing creativity and problem-solving skills

#Question Types
  - Select the correct answer related to the execution result or the cause of an error in a given code snippet
  - Choose multiple correct answers related to the behavior or execution result of a code snippet
  - Arrange code snippets or algorithm steps in the correct execution sequence
  - Fill in the blank with appropriate keywords or function names from the given options
  - Select the most suitable algorithm or programming approach for a given problem scenario
  - Match programming terms with their definitions
  - Select the correct combination of code snippets or options to create a valid program
  - Analyze the output of a code snippet and identify the cause of the result
  - Select the explanation that matches the behavior or definition of a code snippet or algorithm
  - Predict the result of modifying or replacing parts of a code snippet
```

**Fig. 2.** Template for Proposed Multiple-Choice Questions

The essay template is designed to assess students' conceptual understanding and problem analysis skills. This template prompts students to analyze problems logically, describe their findings, and propose improvements. The questions combine theoretical and practical aspects, including theoretical problems such as defining and explaining terms, as well as practical problems like identifying and improving issues in code. Easy questions focus on explaining concepts, while more difficult questions require analyzing and optimizing complex code. Key question types include defining and explaining specific programming terms, analyzing and explaining the causes of code errors, and identifying issues in given code while proposing improvements. The essay template helps students deeply understand concepts, apply their knowledge effectively, and strengthen their logical reasoning during the problem-solving process.

*A. Analysis of Results*

Fig. 5 presents examples of questions generated using the proposed multiple-choice template, demonstrating its effectiveness in evaluating students' foundational conceptual understanding and logical reasoning skills. The questions are designed with a clear and concise structure, enabling students to easily grasp the intent of each query. For instance, the question assessing variable naming conventions directly evaluates fundamental programming syntax and minimizes ambiguity through well-defined options. The generated questions are closely aligned with learning objectives, with each question targeting specific programming competencies. The variable naming convention question evaluates basic syntax understanding, the default parameter question encourages deeper comprehension of function design, and the file input/output question focuses on practical data-handling skills. This breadth, ranging from basic to intermediate-level concepts, highlights the template's adaptability to varying levels of learner proficiency.

```
#Directive
You are a [Python] expert responsible for creating exam questions.
Please refer to the template below and create high-quality questions.

#Scope
  - Data types, operators, conditional statements, loops
  - Functions, error handling
  - Iterator, Generator, Coroutine
  - Numpy library

#Basic Information
  - Question type: [short-answer]
  - Programming language: [Python]
  - Number of questions: [10 questions]
  - Difficulty distribution:
    - 1/5: [1 question]
    - 2/5: [1 question]
    - 3/5: [2 questions]
    - 4/5: [3 questions]
    - 5/5: [3 questions]
  - Provide questions, answers, and explanations
  - Specify difficulty for each question

#Question Criteria
  - Single-topic questions
  - Questions integrating multiple topics
  - Questions assessing creativity and problem-solving skills

#Question Types
  - Predict the output of the provided code
  - Fill in the missing parts of the program or complete the entire code
  - Analyze the provided table or graph and complete the missing parts of the program or complete the entire code
  - Generate code to match a given output
  - Write code based on detailed instructions and conditions provided
  - Create complete programs with multiple functions addressing various requirements
```

**Fig. 3.** Template for Proposed Short-Answer Questions

```
#Directive
You are a [Python] expert responsible for creating exam questions.
Please refer to the template and create high-quality questions.

#Scope
  - Shallow-copy, Deep-copy
  - Functions, error handling
  - File input/output(csv, excel)
  - Random library

#Basic Information
  - Question type: [Essay]
  - Programming language: [Python]
  - Number of questions: [5 questions]
  - Difficulty distribution:
    - 1/5: [1 question]
    - 2/5: [1 question]
    - 3/5: [1 question]
    - 4/5: [1 question]
    - 5/5: [1 question]
  - Provide questions, answers, and explanations
  - Specify difficulty for each question

#Question Criteria
  - Single-topic questions
  - Questions integrating multiple topics
  - Questions assessing creativity and problem-solving skills

#Question Types
  - Explain specific programming terms
  - Analyze the cause of errors in provided code and explain
  - Identify problems in given code and propose improvements
```

**Fig. 4.** Template for Proposed Descriptive Questions

## IV. EXPERIMENTS

Additionally, the inclusion of detailed explanations for each question significantly enhances the learning process. These explanations clarify the validity of each choice and help students understand why incorrect answers are flawed. This feedback allows learners to reinforce their conceptual knowledge and address any gaps in understanding. For example, the default parameter question provides a clear explanation of parameter behavior, ensuring a robust grasp of the concept. Lastly, the generated questions exhibit an appropriate diversity in both difficulty and topic coverage. They range from basic syntax and structure questions to more practical applications such as function design and file handling, demonstrating the template's utility for both novice and advanced learners.

**Fig. 5.** Examples of Generated Multiple-Choice Questions and Answers

**Fig. 6.** Examples of Generated Short-Answer Questions and Answers

In conclusion, the multiple-choice template generates questions that are highly effective in terms of clarity, alignment with learning objectives, and feedback provision. This template accommodates diverse learning levels and proves to be a powerful tool for educational assessment.

The questions generated using the proposed short-answer template, as shown in Fig. 6, are designed to evaluate students' problem-solving skills, logical reasoning, and programming proficiency. Each problem is clearly structured, ensuring that students can easily understand and address the tasks presented. For instance, the question requiring the implementation of a function to determine whether a number is even focuses students' efforts on a specific task, while the fill-in-the-blank problem for calculating the sum of a list effectively assesses students' understanding of Python's built-in functions.

The generated problems cover a wide range of topics, from fundamental programming concepts to more advanced subjects. Basic problems, such as writing a function to check even numbers or calculating the sum of a list, help students solidify their understanding of Python syntax and function usage. In contrast, problems involving the NumPy library, such as calculating the sum of an array and explaining its behavior, assess practical data processing and array manipulation skills at an intermediate level. This diversity demonstrates that the template is suitable for learners at various proficiency levels, from beginners to intermediate students. Additionally, the short-answer template encourages students to analyze problems and write appropriate code to solve them. For example, fill-in-the-blank problems test students' ability to identify and apply Python's specific functionalities effectively. The NumPy problem, which requires understanding and explaining the behavior of the axis parameter, provides an opportunity for students to gain hands-on experience with real-world data manipulation techniques.

Each question is accompanied by detailed answers and explanations, which significantly enhance the learning process. For instance, the even-number function problem explains how the modulus operator is used to determine divisibility by two. Similarly, the NumPy array problem provides a detailed explanation of how the axis parameter works to calculate column-wise sums, helping students understand the behavior of multidimensional arrays. These explanations are highly effective in reinforcing concepts and helping students learn from their mistakes. The generated questions also include diverse scenarios such as function implementation, completing code snippets, and predicting the output of code. These scenarios reflect real-world programming challenges and bridge the gap between theoretical learning and practical application. Including libraries such as NumPy ensures that students are exposed to tools and techniques that are relevant in industry and practical contexts.

In conclusion, the short-answer template is a powerful tool designed to help students apply their programming skills and demonstrate their problem-solving abilities. The problems are clear, educationally relevant, and balanced across fundamental and advanced topics. Additionally, the detailed explanations provided for each question make the template an invaluable resource for both assessment and learning, proving its effectiveness as a tool for evaluating and enhancing programming education.

The questions generated using the proposed essay template are designed to deeply evaluate students' conceptual understanding and problem analysis skills, as illustrated in Fig. 7. These questions do not merely seek correct answers but instead prompt students to analyze the root causes of issues, propose solutions, or provide clear explanations of concepts. For instance, the first question asks students to explain the difference between shallow copy and deep copy, along with their respective use cases. This allows students to develop a deeper understanding of complex data structures and memory management. The second question presents a code snippet that triggers a ZeroDivisionError and requires students to propose appropriate exception-handling strategies. Such questions assess students' ability to analyze errors and apply Python's exception-handling mechanisms effectively.

---

**Python Exam Questions: Essay**

---

**Level 1/5 (Basic)**

1. **Question:**
   Explain the difference between shallow copy and deep copy in Python. Why might you use one over the other? Provide examples to illustrate the differences.

**Answer:**
A **shallow copy** creates a new object but does not recursively copy the nested objects. Instead, it references the original nested objects. A **deep copy**, on the other hand, creates a new object and recursively copies all objects it contains, making the copy completely independent of the original.

**Example:**

```python
import copy

original = [[1, 2], [3, 4]]
shallow = copy.copy(original)
deep = copy.deepcopy(original)

shallow[0][0] = 99
print(original)   # [[99, 2], [3, 4]] - Affected because of shared references

deep[0][0] = 88
print(original)   # [[99, 2], [3, 4]] - Not affected due to deep copy
```

**Explanation:**

- Use shallow copy when you want to save memory and are okay with changes in nested structures reflecting in both the copy and the original.
- Use deep copy when you need complete independence between the original and the copy.

**Fig. 7.** Examples of Generated Essay Questions and Answers

The generated questions are structured to guide students toward logically approaching problem-solving and providing detailed explanations. In the first question, students are required to compare the behaviors of shallow copy and deep copy and explain them using example code, ensuring a comprehensive understanding of how data structure copying works in practice. The second question asks students to identify an error and use the try-except construct to propose a solution, enabling them to develop practical debugging and error-handling skills.

Each question includes detailed answers and explanations, which assist instructors in grading and comparing responses. The explanation for the shallow and deep copy question clearly delineates each concept and its behavior, using example code and results to enhance understanding. Similarly, the explanation for the exception-handling question outlines the cause of the error and provides corrected code while clarifying how the try-except construct improves program stability. Such detailed feedback is highly effective in helping students learn from mistakes and reinforce their understanding of key concepts. Essay questions also cover a broad range of topics and strengthen real-world problem-solving abilities. The shallow and deep copy question demands a comprehensive understanding of memory management and data structure behavior, which are critical skills when working with complex data structures in practice. The exception-handling question evaluates students' ability to ensure code robustness, addressing issues commonly encountered in real-world programming.

In conclusion, the essay template is a robust tool for assessing students' ability to analyze questions, logically propose solutions, and apply practical programming skills. The questions are clear and closely aligned with educational objectives, balancing foundational and advanced topics to meet diverse learning goals. Additionally, the detailed explanations provide valuable learning support, demonstrating that the essay template can be effectively utilized as both an assessment and instructional tool.

The previous discussion focused on Python, but the analysis is extended by applying the template to other programming languages, specifically C++ and Java, and comparing the results. The outcomes are shown in Fig. 8 for Java and Fig. 9 for C++. Applying the proposed multiple-choice template to C++ and Java reveals how the unique syntax and educational objectives of each language influence the focus and structure of the generated questions. This demonstrates that the template is not restricted to a specific language and can be flexibly adapted to various programming languages.

---

**Q2.** What is the output of the following code?

```cpp
#include <iostream>
int main() {
    int a = 10, b = 5;
    std::cout << (a > b ? "A is greater" : "B is greater");
    return 0;
}
```

1. A is greater
2. B is greater
3. Compilation error
4. Runtime error

- **Answer:** 1. A is greater
- **Explanation:** The ternary operator checks the condition `a > b` and outputs the corresponding string.

**Fig. 8.** Example of Generated Result Converted to C++

---

**Q5.** What will happen if the following code is executed?

```java
public class Main {
    public static void main(String[] args) {
        int[] numbers = {1, 2, 3};
        System.out.println(numbers[3]);
    }
}
```

1. Prints `0`
2. Prints `null`
3. Throws an `ArrayIndexOutOfBoundsException`
4. Throws a `NullPointerException`

- **Answer:** 3. Throws an `ArrayIndexOutOfBoundsException`
- **Explanation:** The array index `3` is out of bounds since the valid indices are `0, 1, 2`.

**Fig. 9.** Example of Generated Result Converted to Java

C++ questions emphasize the structural characteristics of the language and low-level programming concepts such as memory management. For instance, a question involving the use of the ternary operator to determine the greater of two values evaluates the learner's understanding of operator precedence and conditional substitution. Such questions enable precise code analysis and execution flow understanding, helping learners acquire the technical skills necessary for high-performance systems programming.

Java questions focus on the principles of object-oriented programming (OOP) and runtime error handling. For example, a question addressing the ArrayIndexOutOfBoundsException caused by accessing an out-of-bounds array index effectively evaluates Java's strict runtime rules and exception-handling mechanisms. These questions emphasize structural completeness and code stability, equipping learners with the skills needed to address runtime issues commonly encountered in real-world software development.

The generated questions reflect the unique features and learning objectives of each language, demonstrating the template's versatility. C++ questions demand proficiency in memory management and high-performance programming, targeting intermediate and advanced learners. Java questions address runtime errors and object-oriented design principles, assessing learners' ability to write structured and reliable code.

### B. Survey Evaluations

This study conducted a survey targeting undergraduate students majoring in programming to evaluate the usefulness and educational effectiveness of the proposed problem template. The survey comprised 14 questions, divided into four categories: template functionality (#1-#5), evaluation of problems and answers (#6-#10), naturalness of problems (#11-#12), and learning effectiveness and application intention (#13-#14). The responses were measured on a 5-point Likert scale, ranging from 1 (strongly disagree) to 5 (strongly agree). The survey was conducted online over 20 days, with a total of 46 participants. Respondents were categorized based on programming experience into four groups: less than 1 year, 1-2 years, 2-3 years, and more than 3 years. The categorization of participants based on their programming experience is shown in Figure 10. Table 1 presents the proposed survey questions, and Table 2 provides the mean(M) and standard deviation(SD) categorized by programming experience and overall results.

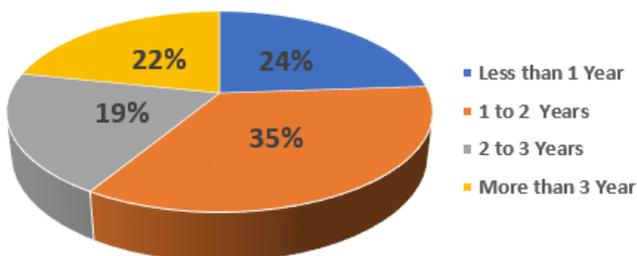

**Fig. 10.** Programming Experience of Survey Participants

TABLE I.    SURVEY QUESTIONS FOR THE PROPOSED METHOD

| | Questions |
|---|---|
| #1 | The ability to set question types (e.g., multiple-choice, short answer, descriptive) is useful. |
| #2 | The ability to specify the programming language (e.g., Python, C++, Java) for problem generation is beneficial. |
| #3 | Differentiating difficulty levels (e.g., 1/5 to 5/5) to tailor problems to the learning level is effective. |
| #4 | The functionality to set the number of problems (e.g., 10 or 20 questions) at once is advantageous. |
| #5 | The capability of generating various types of problems (e.g., error detection, predicting code output, code completion) according to the template instructions is practical. |
| #6 | The generated problems exhibit sufficient diversity. |
| #7 | The explanations of the problem statements are clear and easy to understand. |
| #8 | The programming knowledge incorporated into the problems is appropriate and well-reflected. |
| #9 | The answers (or solutions) are logically valid and free of errors. |
| #10 | The problems are well-suited for assessing the learned material. |
| #11 | The generated problems are comparable in quality to those created by humans. |
| #12 | The wording and expressions in the problems are natural and not awkward. |
| #13 | Using the template to encounter problems of varying difficulty and formats can contribute to improving grades in relevant courses. |
| #14 | Utilizing the generated content for future exams (e.g., midterms or finals) is considered a positive approach. |

Overall, all groups gave positive evaluations of the AI-based problem generation system; however, there were clear differences across specific items. Particularly noteworthy was that the less than one year group consistently showed large standard deviations, indicating a lack of consistency in responses. This suggests that beginners' lack of experience leads to greater variability and instability in their perceptions.

#1 (Ability to set problem types) recorded an overall average of 4.30 (0.78), indicating relatively positive evaluations. Looking at group-level results, the 2-3 years group rated it the highest at 4.67 (0.50), recognizing that selecting problem types meaningfully contributed to their learning process. The 1-2 years group also rated it positively at 4.44 (0.63). In contrast, the less than one year group gave the lowest score of 3.91 (1.14), with the largest standard deviation, showing that beginners with limited learning experience had difficulty strategically utilizing problem type distinctions, and their opinions were widely dispersed. This can be explained by constructivist learning theory: when prior experience is lacking, learners cannot fully grasp the meaning of new tools.

# 2 (Ability to specify programming languages) received the highest overall rating, with an average of 4.53 (0.62). The 2-3 years group gave the strongest evaluation at 4.89 (0.33), reflecting the belief that language-specific problem provision directly contributes to achieving learning goals. The 1-2 years group also rated it highly at 4.60 (0.51), and the 3 years or more group followed with a stable 4.40 (0.52). Although the

less than one year group rated it lower at 4.27 (0.79), it was still positive. These results are aligned with self-directed learning theory, which emphasizes that learners are more motivated and satisfied when choosing resources tailored to their own goals.

#3 (Ability to adjust difficulty level) had an overall average of 4.07 (0.81), which was somewhat lower. Both the less than one year group and the 2-3 years group rated it 4.00 (0.89, 1.00), while the 3 years or more group rated it slightly higher at 4.10 (0.57). The 1-2 years group was relatively more positive at 4.13 (0.72). This suggests that beginners may find difficulty distinctions ambiguous, while advanced learners place greater importance on the depth and authenticity of problems rather than predefined levels. From the perspective of cognitive load theory, novices may perceive unnecessary difficulty distinctions as extraneous load.

#4 (Ability to set the number of problems) was rated highly with an overall average of 4.43 (0.65). The 2-3 years group gave the highest rating of 4.56 (0.53), followed by the 1-2 years group at 4.50 (0.63). The 3 years or more group rated it 4.30 (0.48), and the less than one year group 4.36 (0.81). All groups evaluated this item positively. This finding can be linked to self-regulated learning theory, as the ability to control workload serves as an important tool for learner autonomy, particularly in intermediate stages.

#5 (Ability to generate diverse problem types) had an overall average of 4.44 (0.69). The 1-2 years group rated it the highest at 4.69 (0.48), and the 3 years or more group also rated it highly at 4.67 (0.50). The 2-3 years group rated it 4.56 (0.53). In contrast, the less than one year group rated it the lowest at 3.82 (0.98), with the largest standard deviation. This indicates that beginners often perceive diversity as complexity, while intermediate and advanced learners value diversity as a factor for sustaining motivation and promoting knowledge transfer.

#6 (Diversity of problems) had an overall average of 4.09 (0.72). The 3 years or more group rated it highest at 4.30 (0.48), followed by the 2-3 years group at 4.22 (0.67), and the 1-2 years group at 4.06 (0.70). The less than one year group gave the lowest score of 3.82 (0.87). This shows that beginners prioritize comprehensibility over diversity, consistent with cognitive load theory, where learners aim to reduce extraneous load.

#7 (Clarity of problem statements) had an overall average of 4.07 (0.74). The 2-3 years group rated it the highest at 4.33 (0.50), followed by the 3 years or more group at 4.10 (0.57),

the 1-2 years group at 4.00 (0.74), and the less than one year group at 3.91 (0.83). This suggests that beginners, who already perceive problems as difficult, are particularly sensitive to unclear statements.

#8 (Appropriateness of programming knowledge) had an overall average of 4.28 (0.66). The 3 years or more group gave the highest score of 4.50 (0.53), followed by the 2-3 years group at 4.33 (0.50) and the 1-2 years group at 4.31 (0.70). The less than one year group rated it lowest at 4.00 (0.77). This indicates that advanced learners are more attentive to whether problems align with actual programming knowledge and curricula.

#9 (Logical validity of answers/solutions) had an overall average of 4.09 (0.75). Interestingly, the less than one year group gave a relatively high rating of 4.27 (0.90), likely due to limited ability to critically evaluate logical errors. In contrast, the 1-2 years group gave the lowest score of 3.81 (0.81), reflecting that learners at this stage are more attentive to accuracy and consistency in reasoning.

#10 (Suitability for assessment) was rated positively with an overall average of 4.20 (0.59). The 3 years or more group rated it highest at 4.40 (0.52), followed by the 2-3 years group at 4.22 (0.67). The less than one year group gave the lowest rating of 3.91 (0.83), suggesting that advanced learners recognize the assessment utility of problems more than beginners.

#11 (Comparison with human-created problems) recorded the lowest overall rating, with an average of 3.89 (0.61). Notably, the less than one year group gave the lowest score of 3.55 (0.69). This may not simply be due to lack of experience, but rather because beginners lack the evaluative framework to compare AI-generated and human-created problems, perceiving the former as unfamiliar or incomplete. The 1-2 years group rated it more positively at 4.19 (0.64), while the 2-3 years group and 3 years or more group rated it lower again at 3.89 (0.60) and 3.80 (0.42), respectively. This suggests that intermediate learners tend to take a more practical view, while advanced learners evaluate more critically, emphasizing creativity, contextual appropriateness, and completeness.

#12 (Naturalness of wording and expressions) was rated highly, with an overall average of 4.35 (0.66). The 3 years or more group gave the highest rating of 4.50 (0.53), and even the less than one year group responded positively with 4.27 (0.79). Natural and intuitive expressions make problems more accessible for novices and more professional for advanced learners.

TABLE II.    RESULTS OF THE PROPOSED METHOD SURVEY (DETAILED SCORES BY PROGRAMMING EXPERIENCE AND OVERALL SCORES)

|  | #1 | #2 | #3 | #4 | #5 | #6 | #7 | #8 | #9 | #10 | #11 | #12 | #13 | #14 |
|---|---|---|---|---|---|---|---|---|---|---|---|---|---|---|
| 1 Year ↓ | 3.91 (1.14) | 4.27 (0.79) | 4.00 (0.89) | 4.36 (0.81) | 3.82 (0.98) | 3.82 (0.87) | 3.91 (0.83) | 4.00 (0.77) | 4.27 (0.90) | 3.91 (0.83) | 3.55 (0.69) | 4.27 (0.79) | 4.09 (0.83) | 3.73 (0.65) |
| 1 to 2 Years | 4.44 (0.63) | 4.60 (0.51) | 4.13 (0.72) | 4.50 (0.63) | 4.69 (0.48) | 4.06 (0.70) | 4.00 (0.74) | 4.31 (0.70) | 3.81 (0.81) | 4.38 (0.63) | 4.19 (0.64) | 4.31 (0.60) | 4.44 (0.63) | 4.31 (0.70) |
| 2 to 3Years | 4.67 (0.50) | 4.89 (0.33) | 4.00 (1.00) | 4.56 (0.53) | 4.56 (0.53) | 4.22 (0.67) | 4.33 (0.50) | 4.33 (0.50) | 4.22 (0.67) | 4.00 (0.71) | 3.89 (0.60) | 4.33 (0.50) | 4.67 (0.50) | 4.11 (0.60) |
| 3 Year ↑ | 4.20 (0.63) | 4.40 (0.52) | 4.10 (0.57) | 4.30 (0.48) | 4.67 (0.50) | 4.30 (0.48) | 4.10 (0.57) | 4.50 (0.53) | 4.20 (0.42) | 4.40 (0.52) | 3.80 (0.42) | 4.50 (0.53) | 4.20 (0.42) | 4.20 (0.42) |
| All | 4.30 (0.78) | 4.53 (0.62) | 4.07 (0.81) | 4.43 (0.65) | 4.44 (0.69) | 4.09 (0.72) | 4.07 (0.74) | 4.28 (0.66) | 4.09 (0.75) | 4.20 (0.59) | 3.89 (0.61) | 4.35 (0.66) | 4.35 (0.62) | 4.11 (0.64) |

#13 (Potential for enhancing learning outcomes) had a high overall average of 4.35 (0.62). The 2-3 years group gave the highest score of 4.67 (0.50), indicating that intermediate learners feel the greatest benefits from template-based problem generation. In contrast, the less than one year group gave a lower rating of 4.09 (0.83), showing that beginners are less able to fully perceive its potential benefits.

#14 (Applicability in exams) recorded an overall average of 4.11 (0.64). The 1-2 years group gave the highest rating of 4.31 (0.70), followed by the 3 years or more group at 4.20 (0.42) and the 2-3 years group at 4.11 (0.60). The less than one year group rated it lowest at 3.73 (0.65), reflecting their lack of exam experience and limited awareness of assessment contexts.

In summary, while all items were rated positively overall, the less than one year group consistently gave lower averages with larger standard deviations, showing greater variability in perceptions. This suggests that beginners, due to lack of experience, struggle to consistently recognize the utility of system functions. In contrast, the 1-2 years and 2-3 years groups provided higher and more consistent ratings, particularly emphasizing programming language specification, workload control, and learning effectiveness. These features align closely with self-directed learning and reflect the growing ability of learners to connect system functions with their personal learning goals.

*C. Discussions*

The proliferation of generative AI is fundamentally reshaping the paradigm of programming education. In the past, learners primarily focused on acquiring the syntax and grammar of programming languages and writing code manually. Today, however, learners increasingly interact with language models such as ChatGPT to accelerate the problem-solving process or delegate code generation. Within this transformation, prompt engineering is no longer a mere technical tool but emerges as a core learning competency that enables learners to design and regulate their own learning process through interaction with AI.

From an educational perspective, the contribution of the proposed approach can be understood on multiple levels. First, prompt engineering requires learners to define problems, specify context, and formulate conditions for the AI system. This process constitutes a metacognitive activity, in which learners externalize and structure their own thinking. In doing so, learners are not only searching for answers but also analyzing problem situations and clarifying learning objectives. Such practices provide concrete implementation of constructivist learning theory, which emphasizes the active construction of knowledge by learners. Second, the proposed template was explicitly designed with Bloom's Taxonomy in mind, encouraging learners to move beyond knowledge recall and comprehension to engage in analysis, evaluation, and creative problem-solving. By requiring learners to verbalize their reasoning, articulate solutions, and explore multiple approaches, the automatically generated questions cultivate higher-order thinking skills. This represents a pedagogical shift in programming education from rote memorization toward problem-solving-centered learning.

Third, prompt engineering-based question design also holds implications for instructors. Traditionally, constructing programming exam questions demanded significant time and expertise, often reducing the efficiency of instructional practice. The proposed approach, however, allows instructors to generate diverse questions efficiently while aligning them with specific learning objectives, thereby augmenting instructional design capacity. Moreover, the generated questions maintain consistent structure and coherence, which can enhance both reliability and validity in assessment practices. Fourth, the proposed template reframes assessment by moving beyond the traditional question of "what is the correct answer" to emphasize how learners interpret a problem, construct logical reasoning, and present their solutions clearly. This shift supports learners in applying programming knowledge to authentic contexts and articulating their cognitive processes, thereby facilitating the transfer of learning. In this sense, AI-generated exam questions serve not only as assessment tools but also as catalysts for deeper learning.

In conclusion, the proposed template demonstrates a clear educational contribution within the context of generative AI-driven programming education. Specifically, it promotes learners' metacognitive reflection, systematically develops higher-order thinking skills, supports instructors' instructional design and assessment capacity, and strengthens learners ability to apply knowledge in authentic contexts. The value of this study therefore lies not solely in technical innovation but in establishing a pedagogical turning point that reorients programming education toward learner-centered and problem-solving-oriented practices.

## V. CONCLUSION

This study proposed a template utilizing prompt engineering techniques for the effective generation of exam questions and answers in programming education, analyzing its application across various programming languages. The findings demonstrate that the proposed template is a robust tool capable of comprehensively evaluating learners' conceptual understanding, problem-solving abilities, and logical reasoning through multiple-choice, short-answer, and essay-style questions. The generated questions were designed to reflect the unique syntactic features and learning objectives of different programming languages, confirming the template's flexibility and adaptability to diverse languages without being confined to any specific one. The proposed template covers a wide range of programming language concepts, with question types systematically designed to assess learners' understanding and ability to apply these concepts. The questions extend beyond requiring simple answers, encouraging learners to analyze problems, propose appropriate solutions, or clearly articulate concepts. This approach fosters a deeper understanding of programming concepts and practical application skills among students.

As a result, the automatically generated questions were found to contribute to achieving actual educational objectives rather than simply offering superficial diversity in format. In a pilot analysis, learners reported that the generated questions helped them clarify conceptual understanding and systematize their problem-solving processes. This indicates that the proposed template is not merely a technical automation tool but a mechanism with tangible impact on strengthening the learning process. Nevertheless, this study has certain limitations. Long-term datasets that trace learning outcomes remain insufficient, and further rigorous methodologies and extended validation are necessary to comprehensively evaluate the effects of the template on sustained learning and actual performance improvement.

Future work should explore the following directions. First, developing an automated system for systematically adjusting question difficulty would enable the generation of questions tailored to learners' proficiency levels, enhancing the efficiency of difficulty management. Second, improving the design of the template to support question generation optimized for diverse programming languages and learning objectives is crucial. Incorporating automated prompt optimization techniques can significantly extend the template's applicability and effectiveness. Third, the template should be extended beyond programming education to other disciplines, such as mathematics, physics, and data science, where prompt-based question generation could serve as an effective learning and assessment tool. Fourth, conducting user studies involving both students and instructors is essential to evaluate the practical educational impact of the template. Such studies could measure satisfaction and learning results, providing insights for further refinement and enhancement.

In conclusion, the proposed template demonstrates significant potential as an effective learning and assessment tool in programming education. It combines clarity and educational alignment, contributing to the deepening of students' problem-solving skills and logical reasoning. Future research should aim to further enhance the template's versatility and practicality, thereby advancing the overall quality of programming education.